# Sphere-to-cylinder transition in hierarchical electrostatic complexes

Jean-François Berret

*Matière et Systèmes Complexes, UMR 7057 CNRS Université Denis Diderot Paris-VII, Bâtiment Condorcet, 10 rue Alice Domon et Léonie Duquet, 75205 Paris, France*

RECEIVED DATE (Tuesday 31 March 2009); Author E-mail Address: jean-francois.berret@univ-paris-diderot.fr

**Abstract** : We report the formation of colloidal complexes resulting from the electrostatic co-assembly between anionic surfactants and cationic polyelectrolytes or block copolymers. Combining light and x-ray scattering experiments with cryogenic transmission and optical microscopy, we emphasize a feature rarely addressed in the formation of the electrostatic complexes, namely the role of the mixing concentration on the microstructure. At low mixing concentration, electrostatic complexes made from cationic-neutral copolymers and alkyl sulfate surfactants exhibit spherical core-shell microstructures. With increasing concentration, the complexes undergo a sphere-to-cylinder transition, yielding elongated aggregates with diameter 50 nm and length up to several hundreds of nanometers. From the comparison between homo- and diblock polymer phase behaviors, it is suggested that the unidimensional growth is driven by the ability of the surfactant to self-assemble into cylindrical micelles, and in particular when these surfactants are complexed with oppositely charged polymers.

## I – Introduction

Controlled electrostatic interactions between colloidal species at the nanoscale have attracted much interest in the recent years [1-4]. It is expected that such interactions could direct the association between components of different nature such as organic and mineral or synthetic and biological, as well as to build large supracolloidal structures with multimodal attributes. Because bottom-up strategies of nanofabrication cover a large spectrum of sizes, typically between 1 nm and 1 µm, hybrid structures appeared as promising tools for applications dealing with drug delivery, catalysis, functionalization, biomedecine, sensors and membranes. The requirements for electrostatics based assemblies are a high density of charges on the surface of nanocolloids or along the backbone of polymers. A second requirement is that the charges of both constituents are opposite. In contrast to self-assembly which deals with single component systems such as surfactants or amphiphilic copolymers, when two constituents of different nature are used, the term co-assembly is preferred [5,6].

In the past years, oppositely charged surfactant and polyelectrolytes (PE) were extensively studied with respect to their bulk [7-15] and surface [10,16-18] properties. Among the cationic PEs considered, poly(diallyl dimethylammonium chloride), poly(trimethylammonium methyl acrylate) and poly(ethyleneimine) were looked at carefully and the complexation schemes were elaborated in combination with alkyl sulfate surfactants, and in particular with sodium dodecyl sulfate (SDS). In the dilute range of concentrations, the surfactant/PE/water phase diagrams were found to display a liquid-solid phase separation. When centrifuged, the solutions exhibited a coexistence of two phases, the bottom phase being a precipitate containing most organic species. The knowledge of the precipitate structures at local and intermediate scales is of importance for the understanding of the complexation mechanisms. Surfactant/PE precipitates were investigated repeatedly during the past years by means of scattering experiments and it was found that they were liquid-crystalline mesophases. Mesophases with cubic, hexagonal or lamellar symmetries were disclosed, depending on whether the surfactant micelles (formed through complexation with the PEs) were spherical, cylindrical or planar. Through a comprehensive study of alkyl sulfate surfactants (with alkyl chains comprising 8 to 14 $CH_2$ groups), Zhou *et al.* [9,11] and Sokolov *et al.* [8] have established that the nature of the mesoscopic long-range order depended on the length of the alkyl chain, and to a lesser extent on the nature of the polyelectrolyte [11]. Concerning SDS however, all reports known to us agreed that the precipitates exhibited hexagonal order, with SDS cylindrical micelles packed according to a long-range orientational and translational hexagonal symmetry [7-9,12,14,15].

As for their interfacial properties, it was shown that surfactant/PE complexes deposited spontaneously onto mica and poly(styrene) substrates in a large amount, provided that the charge ratio between surfactant and polymers was adjusted below the stoichiometry [16-18]. Many studies also explored the role of the adjunct of a neutral block onto the polyelectrolyte chains, yielding either a block [14,19], graft or cross-linked [20-22] copolymers. With block copolymers, the phase separation described above was found to be replaced by the spontaneous formation of core-shell colloids. The core-shell microstructure of the colloids was ascertained using small-angle neutron and x-ray scattering [23,14,19]. In SDS/PTEA$_{11K}$-*b*-PAM$_{30K}$, a system that is described in detail below, these two techniques allowed to depict the core as an assembly of spherical micelles linked by the polyelectrolyte blocks. The identification was made possible through the detection of a broad and unresolved structure peak at high wave-vector, at q* ~ 0.16 Å$^{-1}$. Investigations using anionically charged polymers and copolymers were also performed in parallel and provided similar results in terms of phase separation [24-26,9,27-32,14,33] and controlled colloidal complexation [34-36,30,31,14,37,38], respectively .

In the present paper, we emphasize a feature rarely addressed in the formation of the electrostatic complexes, namely the role of the mixing concentration on the microstructure. The majority of electrostatics based complexes, and in particular those produced with block copolymers were found so far to be spherical [39,34,40,41,30,31,23,42,43,5,44,6]. Here, however it is shown that complexes can exhibit cylindrical morphologies, as revealed by SAXS and cryo-TEM experiments. At low concentration, the aggregates are spherical, whereas at higher concentrations, the structure elongates to reach size of several hundreds of nanometers. The use of two alkyl sulfate surfactants differing in aliphatic lengths (with 10 and 12 carbon atoms) aimed to illustrate the generality of the present approach. From the comparison between homopolyelectrolyte and diblock copolymers, it is suggested that the unidimensional growth is driven by the ability of the surfactant to self-assemble into cylindrical micelles.

## II – Experimental

II.1 – Material, Polymer Characterization and Sample Preparation

Poly(trimethylammonium ethylacrylate methyl sulfate) homopolyelectrolytes and poly(trimethylammonium ethylacrylate methyl sulfate)-b-poly(acrylamide) block copolymers were synthesized by MADIX® controlled radical polymerization [45]. The polymers used were abbreviated PTEA$_{11K}$ and PTEA$_{11K}$-b-PAM$_{30K}$ respectively, where the indices indicate the molecular weights targeted by the synthesis. PTEA11K was an aliquot removed during the synthesis of the diblock. In aqueous solutions and neutral pH conditions, the chains are well dispersed and in the state of unimers. The weight-averaged molecular weights and the hydrodynamic diameter of the polymers were determined by static light scattering experiments ($M_W$ = 35000 ± 2000 g mol$^{-1}$ and $D_H$ = 11 nm [19]). The polydispersity index was estimated by size exclusion chromatography at 1.6.

Sodium decyl sulfate (SdS) and sodium dodecyl sulfate (SDS) were purchased from Sigma and used without further purification. The critical micellar concentrations for these two systems are 0.86 wt. % (33 mmol l$^{-1}$) for SdS [46] and 0.21 wt. % (8.3 mmol l$^{-1}$) for SDS [47]. Above the cmc, the surfactant aggregates into spherical micelles, the aggregation numbers being of the order of 50 [19]. Also studied with respect to their phase behavior, both surfactants exhibit hexagonal mesophases at concentrations around 40 wt. % [48-50]. Note that in order to observe hexagonal mesophases from SDS solutions, the temperature had to be above the Krafft temperature, *i.e.* 35 °C at these concentrations [49].

*Sample Preparation* : The samples were prepared by adding the surfactant solution to the polymer solution in a one-shot process. Both stock solutions were free of added salt and made at the same weight concentration c and pH (pH 7). The relative amount of each component was monitored by the stoichiometric ratio for chargeable groups Z. The charge ratio reads $Z = [S]/(n_{PE} \times [P])$, where [S] and [P] are the molar surfactant and polymer concentrations and $n_{PE}$ is the degree of polymerization of the polyelectrolyte block ($n_{PE}$ = 41 for PTEA$_{11K}$). Z = 1 describes an isoelectric solution, that is a solution characterized by the same number densities of positive and negative chargeable ions. In the mixed solutions, the surfactant and polymer concentrations $c_S$ and $c_P$ read $c_S = c Z / (z_0 + Z)$ and $c_P = c z_0 / (z_0 + Z)$ with $z_0 = M_w^P / n_{PE} M_w^S$. The description of the mixed solutions in terms of c and Z allowed us to compare polymers of different molecular weights, such as homopolyelectrolyte and cationic-neutral block copolymers. In this work, we focus on surfactant/polymer systems at Z = 1 and c = 0.1 - 30 wt. %. Finally, as shown recently using light scattering measurements in the very dilute regime of concentrations (10$^{-4}$ – 1 wt. %), the surfactant-based complexes SDS/PTEA$_{11K}$-b-PAM$_{30K}$ were shown to exhibit a critical association concentration (cac), estimated at $c_{cac}$ = 5.0×10$^{-3}$ wt. % [51,52].

II.2 – Experimental Techniques and Data Analysis
*Dynamic Light Scattering*
Dynamic light scattering was performed on a Brookhaven spectrometer (BI-9000AT autocorrelator) for measurements of the collective diffusion constant D(c) of the SDS/PTEA$_{11K}$-b-PAM$_{30K}$ dispersions. The autocorrelation functions of the scattered light were determined for various wave-vectors $q = (4\pi n/\lambda)\sin(\theta/2)$ and at different concentrations in the range 0.1 – 1 wt. %. In the above expression, n is the refractive index of the solution, λ the wavelength of the incident beam (λ = 488 nm) and θ the scattering angle. For the solutions put under scrutiny, the autocorrelation functions were all characterized by a single relaxation mode, that was associated to the hydrodynamic size of the colloids through the Stokes-Einstein relationship, $D_H = k_B T / 3\pi\eta_0 D(c\rightarrow 0)$, where $k_B$ is the Boltzmann constant, T the temperature (T = 298 K) and η0 the solvent viscosity ($\eta_0$ = 0.89×10$^{-3}$ Pa s). The collective diffusion coefficients D(c) were found to be concentration independent and varying as q$^2$, as expected from diffusive concentration fluctuations. The autocorrelation functions were analyzed by the cumulant technique and the quadratic diffusion coefficient was considered for calculating $D_H$.

*Cryogenic transmission electron microscopy (cryo-TEM)*
Cryo-TEM experiments were performed on surfactant/polymer dispersions made at concentration c = 0.2 wt. %. For cryo-TEM experiments, a drop of the solution was put on a TEM-grid covered by a 100 nm-thick polymer perforated membrane. The drop was blotted with filter paper and the grid was quenched rapidly in liquid ethane so as to avoid the crystallization of the aqueous phase. The membrane was then transferred into the vacuum column of a TEM-microscope (JEOL 1200 EX operating at 120 kV) where it was maintained at liquid nitrogen temperature, allowing a 20000×magnification.

*Optical Microscopy*
Optical microscopy visualizations were performed with a Leica DM/LB direct microscope working with a pair of analyzer and polarizer apart from the glass cover plate. The textures of the SDS/PTEA$_{11K}$ precipitates were observed using 60× and 100×objectives, allowing to access a field of view between 20 and 100 μm .

| species | chemical formula | $M_W$ g mol$^{-1}$ | $v_0$ (Å$^3$) | $\rho_X$ (e Å$^{-3}$) |
|---|---|---|---|---|
| decyl sulfate | (C$_{10}$H$_{21}$)-OSO$_3^-$ | 260 | 359 | 0.36 |
| dodecyl sulfate | (C$_{12}$H$_{25}$)-OSO$_3^-$ | 288 | 412 | 0.35 |
| trimethylammonium ethylacrylate | CH$_2$CH-COO(C$_2$H$_4$)-N$^+$(CH$_3$)$_3$ | 158 | 210 | 0.41 |
| acrylamide | CH$_2$CH-CONH$_2$ | 71 | 105 | 0.36 |
| water | H$_2$O | 18 | 30 | 0.33 |

**Table I** : Molecular weight $M_W$ (g mol$^{-1}$), molecular volume $v_0$ (Å$^3$) and electronic density $\rho_X$ (e Å$^{-3}$) of the chemical species studied in this work.

*Small-Angle X-ray Scattering*
Small-angle x-rays scattering (SAXS) runs were performed at the Brookhaven National Laboratory (Brookhaven, USA) on the X21 beam line. The scattering configuration used an incoming beam operating at the wavelength of 1.76 Å and the sample-detector distance of 1 m. With the detector in the off-center position, the accessible q-range was 0.01 Å$^{-1}$ – 0.3 Å$^{-1}$, with a resolution $\Delta q_{Res}/q$ of 1.25 %, where $\Delta q_{Res}$ denotes the full width at half maximum (FWHM) of the instrumental resolution function. The data were treated so as to remove the contribution of the 1.5 mm-glass capillaries and that of the solvent. The q-scale was calibrated against silver behenate powder. Table I lists the molecular weights ($M_W$), molecular volumes ($v_0$) and electronic densities $\rho_X$ of the surfactants and monomers studied by SAXS. In this work, as shown in Table I, all chemical species involved in the complexation contributed to the scattering cross-sections. The weak contrast with respect to $H_2O$ was compensated by the high flux of the synchrotron line.

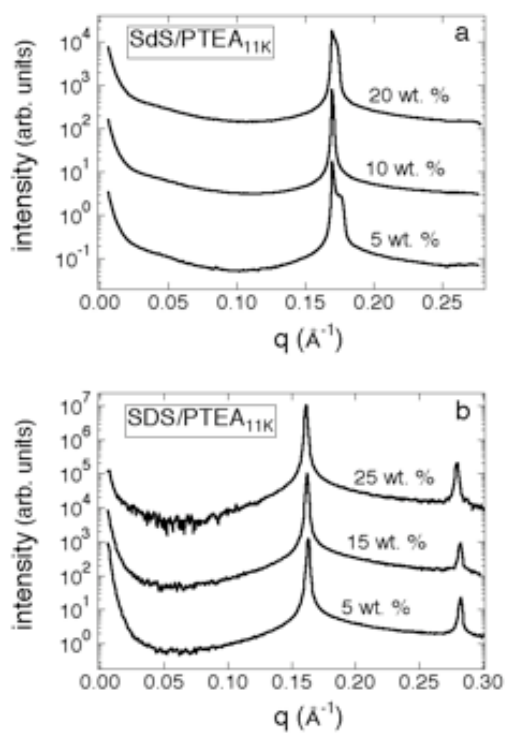

*Figure 1 :* X-ray diffraction intensities for SdS/PTEA$_{11K}$ (a) and SDS/PTEA$_{11K}$ (b) precipitates obtained at mixing concentrations comprised between c = 5 and 25 wt. % and charge ratio Z = 1. Assuming an hexagonal structure for the two systems, the average distance between the cylindrical aggregates was estimated at 4.29 and 4.50 nm, respectively.

## III – Results and discussion

### III.1 – Electrostatic Complexation using Homopolyelectrolytes

The mixing of homopolyelectrolyte PTEA11K and sodium alkyl sulfate surfactants at 1:1 charge ratios resulted in a liquid-solid phase separation over a broad concentration range [14]. The solid phase appeared as a whitish and gel-like precipitate which was separated from the supernatant by centrifugation. In the present work, the precipitates were investigated by small-angle x-ray scattering and by optical microscopy.

Figs. 1a and 1b show the SAXS diffraction intensities obtained from SdS/PTEA11K and from SDS/PTEA11K precipitates, respectively. Concentrations range from 5 and 25 wt. %. For SdS/PTEA11K, only one Bragg reflection was observed using the current configuration. Located at 0.169 Å$^{-1}$, its position remained concentration independent (Table II). Note that at 5 wt. %, the peak was found to broaden, probably due to the coexistence of two close structures. The unique Bragg peak did not allow to retrieve the structure of the SdS/PTEA11K dense phase. By analogy with the work of Zhou et al. [11] on similar systems however, a structure with hexagonal symmetry was anticipated.

| Surf. | polymer | $q_1$ or $q^*$ Å$^{-1}$ | $q_2$ Å$^{-1}$ | Figure |
|---|---|---|---|---|
| SdS | PTEA$_{11K}$ | 0.169 | - | 1a |
| SdS | PTEA$_{11K}$-b-PAM$_{30K}$ | 0.170 | - | 6a |
| SDS | PTEA$_{11K}$ | 0.161 | 0.279 | 1b |
| SDS | PTEA$_{11K}$-b-PAM$_{30K}$ | 0.162 | 0.276 | 6b |

*Table II :* List of wave-vectors, $q^*$, $q_1$ and $q_2$ obtained for the structure and Bragg peaks of surfactant/polymer precipitates and dispersions.

For SDS/PTEA$_{11K}$ precipitates, the occurrence of Bragg peaks at $q_1$ = 0.161 Å$^{-1}$ and $q_2$ = 0.279 Å$^{-1}$ *i.e.* with a spacing ratio √3 suggests an hexagonal order (Table II). The SDS molecules are assumed to associate into elongated cylindrical micelles, in a structure that maintains both translational and orientational long-range orders. From the position of the first order peak, we derive the average distance between cylinders of 4.50 nm ( [12]). X-ray data similar to those of Fig. 1 were reported on the system SDS/PCMA, where PCMA stands for poly(trimethylammonium ethylacrylate chloride), an homopolyelectrolyte that differs from PTEA$_{11K}$ only by the nature of the counterion (chloride instead of methyl sulfate). For the SDS/PCMA precipitates, four Bragg peaks compatible with the hexagonal structure were identified by Bergström et al. [12]. The first order was located at $q_1$ = 0.164 Å$^{-1}$, in very good agreement with the present data.

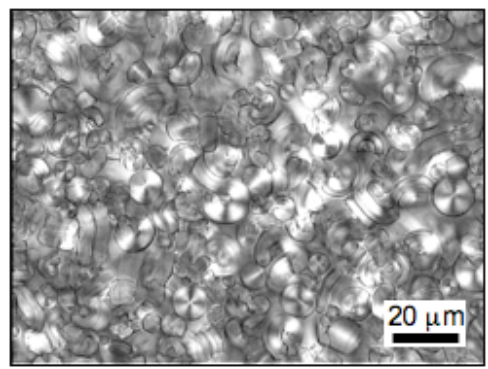

*Figure 2 :* Optical microscopy images of SDS/PTEA$_{11K}$ precipitates (c = 5 wt. %, Z = 1) observed between crossed polarizers (magnification ×60). The sizes of the image is 100×140 µm$^2$. The textures exhibit patterns of few micrometers in size, with bright and dark areas indicative of the spatial changes in the orientations of the SDS cylinders.

Figs. 2 and 3 show microscopy images of the textures formed by SDS/PTEA$_{11K}$ precipitates (c = 5 wt. %) and observed in transmission between crossed polarizers (magnification ×60). The sizes of the images are 100×140 µm$^2$ and 15×30 µm$^2$, respectively. The textures exhibit patterns of few micrometers in size, with spherical and cylindrical morphologies. Each texture displays bright and dark areas indicative of the spatial changes of the refractive index tensor with respect to the polarizer and analyzer positions. The polydomain textures in Fig. 2 are slightly different from the classical fan-like or striated non-geometric textures characteristic of hexagonal phases [53-55]. In lyotropic systems, fan-shaped textures result from a thermomechanical undulation of the micellar columns induced at the isotropic-to-hexagonal transition by decreasing the temperature [54]. In Fig. 3, a uniform domain was isolated and investigated at different angles with respect to the crossed polarizers. Images at angles 0, 30, 60 and 90° are shown together with the positions of the crossed polarizers.

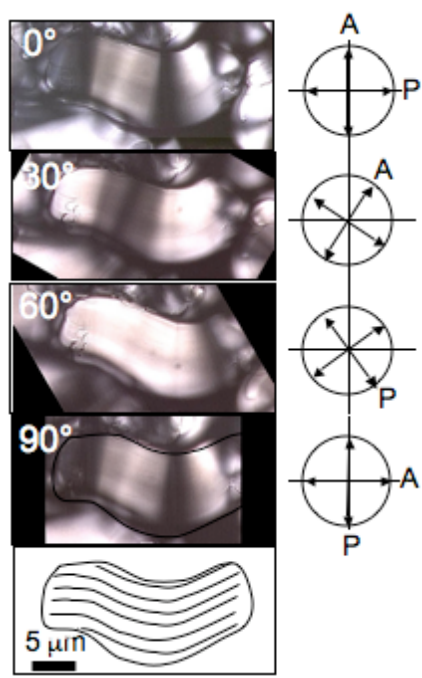

*Figure 3* : A textured pattern of Fig. 2 was isolated and investigated at different angles (0°, 30°, 60° and 90°) with respect to the crossed polarizers. The positions of the polarizer-analyzer are indicated in the right-hand panels. The lower panel illustrates a possible set of orientations for the SDS cylinders within the texture.

The lower panel in the figure illustrates a map of the orientations within the texture. By analogy with the observations of hexagonal mesophases, it is assumed that these orientations are those of the surfactant cylinders forming the hexagonal structure. In a system close to the one studied here, using poly(diallyldimethylammonium chloride) polyelectrolyte and SDS, Nizri and coworkers [15] have shown by cryo-TEM the existence of large domains (200 nm – 1 µm) that had typically the same morphology as the textures in Figs. 2 and 3. With cryo-TEM however, the inside of the domains could be resolved and it was shown to display long (> 1 µm) SDS cylinders aligned parallel the main axis and packed into an hexagonal phase [15]. The distance between the cylinders within the domains was estimated at 4.24 nm, *i.e.* close to that of the present work (4.50 nm for SDS/PTEA$_{11K}$).

III.2 – Electrostatic Complexation using Block Copolymers

When the former homopolyelectrolyte chains were replaced by block copolymers, the phase separation described previously ceased to occur [34,36,30,31]. The modifications of the liquid-solid phase transition using copolymers were studied by many groups during the last decade. For the SDS/PTEA$_{11K}$-*b*-PAM$_{30K}$ system, it was discussed and analyzed in a series of recent publications [14,19] and therefore only the most salient properties will be recalled here. With block copolymers, at low concentrations, the mixed solutions appeared homogeneous and transparent. Fig. 4 shows the scattering cross-section dσ(q)/dΩ measured by small-angle neutron (open symbols) and x-ray (closed symbols) scattering on the same SDS/PTEA$_{11K}$-*b*-PAM$_{30K}$ solution at c = 1 wt. % and Z = 1. Note that because of the weak neutron scattering contrast of the surfactants and polymers with respect to hydrogenated water, the dispersion of Fig. 4 was prepared with deuterated water.

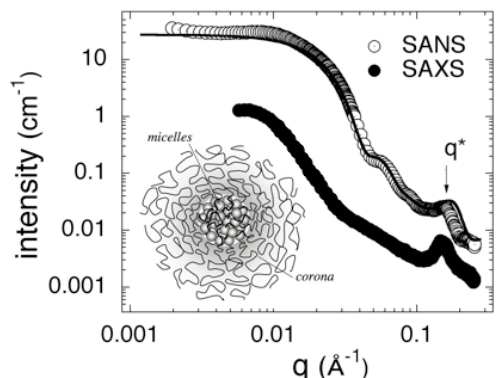

*Figure 4* : Comparison between the SANS and SAXS scattering cross-sections obtained for SDS/PTEA$_{11K}$-*b*-PAM$_{30K}$ complexes at c = 1 wt. % and Z = 1 in D$_2$O [14]. The continuous curve results from Monte Carlo simuations assuming for the assembly the microstructure outlined in the inset. The x-ray intensity is in arbitrary units. Inset : structure assumed for the core-shell aggregates at low mixing concentrations.

In Fig. 4, the SANS and SAXS cross-sections are dominated by two main features : a strong forward scattering with a saturation of the intensity at low wave-vectors and a structure peak at high wave-vectors (arrow). Interestingly, the position of the structure peak at q* observed by the two techniques coincides quite precisely, at q* = 0.152 ± 0.002 Å$^{-1}$. In order to interpret the spectra of Fig. 4 quantitatively, a core-shell structure similar to that of the inset was assumed. In the model, the core was a dense and disordered assembly of surfactant micelles connected by polyelectrolyte blocks, and the structure peak at observed at q* originated from the dense packing of the micelles. Note also that the SDS micelles were supposed to be spherical, with aggregation numbers similar to those of the surfactants at the cmc [19].

The diffuse shell of neutral chains around the core was found to contribute to the scattering in the low q-range. Dynamical light scattering experiments evidenced monodisperse colloids with hydrodynamic sizes D$_H$ = 70 nm, which were attributed to the overall dimension of the core-shell structure.

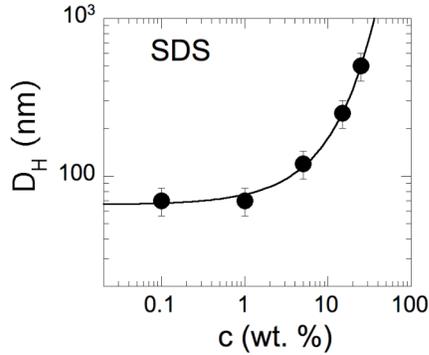

*Figure 5* : Hydrodynamic diameters $D_H(c)$ obtained by dynamic light scattering from SDS/PTEA$_{11K}$-b-PAM$_{30K}$ dispersions prepared at different mixing concentrations c (wt. %), and charge ratio Z = 1. For complexes prepared between 1 wt. % and 20 wt. %, the dispersions were diluted down to 0.1 wt. % for the measurements. Once the colloids were formed, they did not rearrange nor destabilize over time. The continuous line is a guide for the eyes.

The continuous line in Fig. 4 was calculated by Monte Carlo simulations using a hard sphere interaction potential between micelles, and a structure similar to that of the inset. The core diameter was estimated at 17.6 nm and the aggregation number (number of SDS micelles per aggregate) at 71 [19]. The agreement between experimental data and the model was good, especially at high q where the structure factor was correctly reproduced.

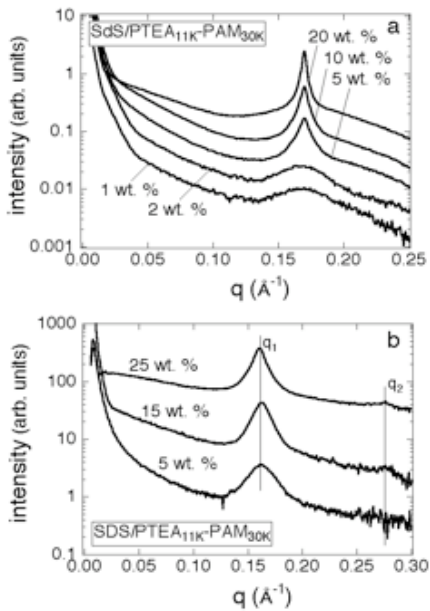

*Figure 6* : SAXS cross-sections obtained from SdS/PTEA$_{11K}$-b-PAM$_{30K}$ (a) and SDS/PTEA$_{11K}$-b-PAM$_{30K}$ (b) dispersions prepared at different mixing concentrations. As c was increased, the structure peak at became narrower and a second order at $q_1/q_2 = \sqrt{3}$ showed up (Fig. 6b). This evolution was interpreted in terms of transition in the surfactant aggregation and microstructure of the colloidal complexes (see text).

In the present paper, we emphasize another feature of the formation of the electrostatic colloidal complexes, namely the role of the mixing concentration on the microstructure of the aggregates. This aspect was rarely addressed in the literature, since most of the systems were studied in the dilute or very dilute regimes of concentration. Fig. 5 displays the hydrodynamic diameters obtained by light scattering from SDS/PTEA$_{11K}$-b-PAM$_{30K}$ dispersions that were prepared at different mixing concentrations (Z = 1). As c increases from 0.1 wt. % to 20 wt. %, $D_H$ was found to grow from 70 nm to 500 nm. For the light scattering experiments, the dispersions were diluted down to 0.1 wt. % so as to avoid absorption and multiple scattering. As long as the concentration remained above the cac, estimated at $5\times10^{-3}$ wt. % for this system [52], the colloids did not rearrange nor they destabilize over time. Such a behavior was also observed using cationic surfactants or inorganic nanoparticles, and it was accounted for by the frozen nature of the surfactant/PE co-assemblies [30,51,52].

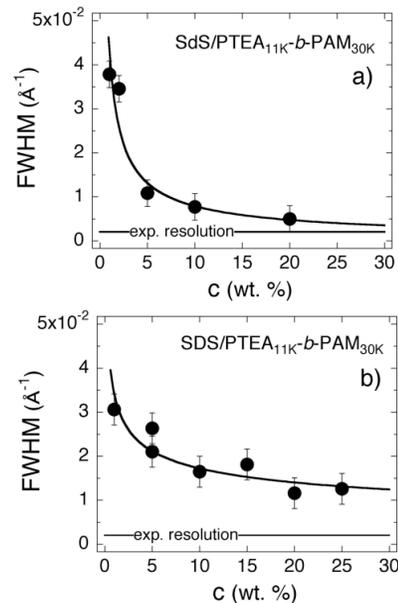

*Figure 7* : FWHM of the structure peak found at high wave-vectors for SdS/PTEA$_{11K}$-b-PAM$_{30K}$ (a) and SDS/PTEA$_{11K}$-b-PAM$_{30K}$ (b) dispersions as a function of the mixing concentrations. The experimental resolution provided by the SAXS spectrometer is shown for comparison. Continuous lines are guides for the eyes.

Figs. 6a and 6b display the SAXS cross-sections obtained from SdS/PTEA$_{11K}$-b-PAM$_{30K}$ and SDS/PTEA$_{11K}$-b-PAM$_{30K}$ dispersions prepared at different mixing concentrations. For the SAXS experiments, the samples were not diluted prior to the measurements. The data are shown in semi logarithmic scales in Figs. 6 in order to emphasize the shape and position of the structure peak observed at high wave-vectors. At low concentrations, the intensity of the structure peak at q* remained weak, with a full width at half maximum broader than the resolution ($\Delta q_{FWHM} \sim 0.04$ Å$^{-1}$ versus $\Delta q_{Res} = 2\times10^{-3}$ Å$^{-1}$). However, as c was increased, the structure peaks became narrower, their intensities increased and a second order at $q_1/q_2 = \sqrt{3}$ showed up. In Figs. 7a and 7b, the FWHMs of the structure peaks for SdS/PTEA$_{11K}$-b-PAM$_{30K}$ and SDS/PTEA$_{11K}$-b-PAM$_{30K}$ systems were plotted as a function of the concentration, and compared to the experimental resolution. At the highest concentrations investigated, the FWHM reached values in the range 0.005 - 0.01 Å$^{-1}$. This evolution was interpreted in terms of microstructural transition of the colloidal complexes. At low c, the surfactants are under the form of spherical micelles closely packed in the core of the colloids, whereas at high c, they co-assemble with the polymers into long cylinders giving rise to the locally hexagonal structure. Using the Debye-Scherrer relationship that relates the full width at half maximum $\Delta q_{FWHM}$ to the spatial extension $\xi_X$ of a crystalline microstructure :

$$\xi_X \sim K \frac{2\pi}{\Delta q_{FWHM} - \Delta q_{Res}}$$

one obtained sizes of 200 nm and 50 nm for SdS and SDS complexes, respectively (in the high concentration limit and using K ~ 0.9 [56,7,57]). For the complexes made from SDS, these values could be compared to those obtained by cryogenic transmission electron microcopy.

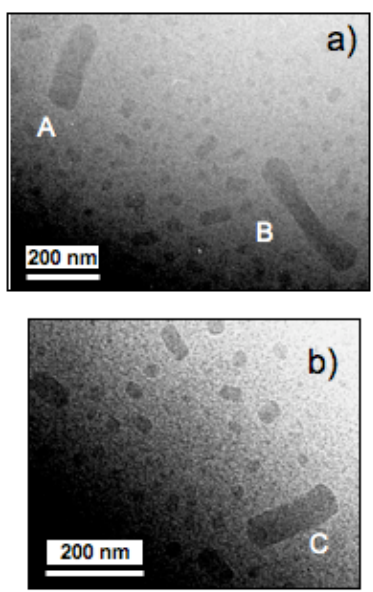

*Figure 8 : Cryo-TEM images of SDS/PTEA$_{11K}$-b-PAM$_{30K}$ electrostatic complexes prepared at c = 5 wt. % and 1:1 charge ratio (magnification ×20000). Aggregates noted A, B (Fig. 8a) and C (Fig. 8b) are cylindrical, with lentghs $L_C$ = 260, 400 and 180 nm and diameters $D_C$ = 90, 70 60 nm, respectively. Note also the rather polydisperse population of aggregates, where small and spherical structures coexist with large and elongated objects.*

Figs. 8 display cryo-TEM images of SDS/PTEA$_{11K}$-b-PAM$_{30K}$ electrostatic complexes prepared at c = 10 wt. % and 1:1 charge ratio. Recorded with a ×20000 magnification, the fields of view are 0.97×0.82 µm$^2$ and 0.57×0.47 µm$^2$ respectively. This magnification allowed us to visualize polydisperse aggregates where small and spherical structures coexist with large and elongated objects. Aggregates noted A and B in Fig. 8a are cylindrical, with lengths $L_C$ = 260 and 400 nm and diameters $D_C$ = 90 and 70 nm. Aggregate C in Fig. 8b exhibits the same morphology with $L_C$ = 180 nm and diameter $D_C$ = 60 nm. We make the hypothesis here that these large aggregates are characterized by a locally hexagonal structure, this structure being induced by the alignment of SDS cylindrical micelles along the major axis of the objects. We also assume that the elongated aggregates are responsible for the narrowing of the Bragg peaks in Fig. 6 b, and the appearance of a second order at $\sqrt{3} q_1$.

For the analysis of the microstructure, it was assumed that the aggregates could be represented by ellipsoids of revolution, and that their 2-dimensional projections were ellipses of major and minor axis $L_C$ and $D_C$, respectively. For the large aggregates of Fig. 8 which morphology was more that of cylinders than ellipses, it was shown that $L_C$ and $D_C$ corresponded quite precisely to the length and diameter of the aggregates. Based on the image analysis of 186 aggregates, the probability distribution functions (pdf) for $L_C$ and $D_C$ were retrieved and plotted in Figs. 9a and 9b, respectively. We found that the pdf for the length was best described by an exponential function, with an average length $L_C^0$ = 27 nm. For the diameter, the pdf was found to be log-normal with a median value $D_C^0$ = 33 nm and a polydispersity $s_D$ = 0.23. These two functions are shown as continuous lines in Figs. 9. Although the average length of the cylinders lies below that of the diameters, it should be recalled that an exponential function is associated to a polydispersity $s_L$ = 2. The large value of the polydispersity in this case is consistent with the existence of elongated aggregates in the 100 nm-range. It should be finally stressed that $D_C^0$ compares well with the size estimates of the hexagonal structures $\xi_X$ as deduced by the Debye-Scherrer relationship. These findings are in agreement with a morphology transition between spherical and cylindrical aggregates.

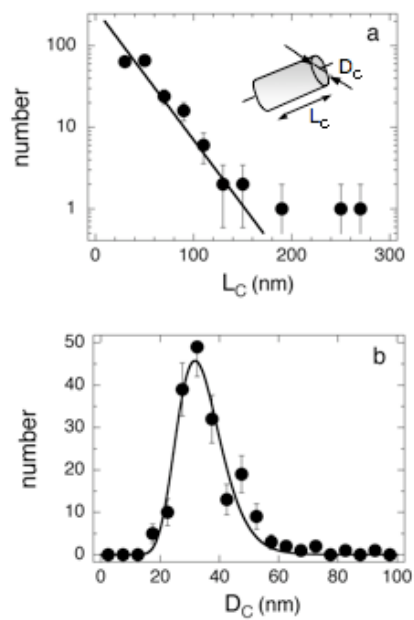

*Figure 9 : Probability distribution function (pdf) for the length $L_C$ and diametrers $D_C$ of the aggregates in Figs. 8. For the analysis of the microstructure, the projected aggregates were assumed to be ellipses of major and minor axis $L_C$ and $D_C$, respectively. $L_C$ was found to be exponentially distributed (average value 27 nm and polydispersity 2), whereas a log-normal function with a median value 33 nm and a polydispersity 0.23 accounted well for the diameters.*

## IV – Concluding remarks

The mixing of oppositely charged surfactants and polyelectrolytes gives rise to compact nanostructures characterized by a crystalline order at the micellar scale. With cationic PEs, sodium decyl and dodecyl sulfate associate into long cylindrical micelles, resulting in an hexagonal mesophase. The results with the homopolyelectrolytes are in agreement with those of the literature on similar compounds. With $PTEA_{11K}$, a phase separation showing the coexistence of the hexagonal structure and the supernatant was clearly evidenced. These conclusions about the local structure were drawn from small-angle x-ray experiments and from a comparison with literature data. The use of two surfactants differing in aliphatic lengths aimed to illustrate the generality of the present results.

It is first interesting to compare the SDS hexagonal phase obtained in the binary system $SDS/H_2O$ and that found using polyelectrolyte complexation schemes. As shown by Kekitcheff [49] for the binary system $SDS/H_2O$, the isotropic-to-hexagonal transition occurred above c = 40 wt. % and for temperature above 35 °C. In this range, the first order peak is located at 0.146 Å$^{-1}$ (T = 36°C and c = 48.5 wt. %), yielding for the intermicellar distance a value of 4.97 nm. With polyelectrolytes such as poly(diallyldimethylammonium chloride), poly(trimethylammonium methyl acrylate) and poly(ethyleneimine), the position of the first Bragg peak was found between 0.161 Å$^{-1}$ and 0.174 Å$^{-1}$, corresponding to a distance between cylinders comprised between 4.2 and 4.5 nm. The distance between cylinders is then smaller with than without polyelectrolytes, an outcome that can be interpreted as the indication of strong intermicellar attractions mediated by the PE chains. Other explanations such as a slight change in the cylinder diameter or the distortion of the hexagonal structure are also plausible [11]. About the value of the structure peaks in the spherical and hexagonal aggregates, it was shown that the wave-vectors q* and $q_1$ remain very close to each other (Table II). This may reflect the fact that during the sphere-to-cylinder transition, as the mixing concentration is increased, the diameters of the spherical and cylindrical SDS micelles remain the same, and so the intermicellar distance.

The main result of the present paper is the observation of submicrometric hierarchical aggregates made from SDS cylinders and packed into a locally hexagonal phase. These elongated mesophases were obtained using block copolymers $PTEA_{11K}$-b-$PAM_{30K}$ instead of homopolyelectrolytes, as previously noted and by increasing the mixing concentration of the constituents. For the majority of electrostatics based co-assembled systems, including systems made from polymers [43,5], proteins [58-60], nanoparticles [61-63] or multivalent counterions [64,65,44], the complexes are generally found to have spherical symmetry. Here, for the first time, we shown that elongated complexes with cylindrical morphology are obtainable as well. SAXS and cryo-TEM were used for the identification of the different microstructures. Interestingly, both techniques have shown consistent results for the sizes of the hierarchical aggregates. Values for the diameter of the order of 50 nm were found, whereas the length can be much larger.

The above findings have several implications. Indirectly, they confirmed that the colloidal complexes prepared and maintained above their cac are out of equilibrium structures [30]. Kinetically frozen, they result from interactions between nanoscale species with opposite charges. This aspect has been documented recently in different kinds of systems, and appears to be an essential attribute for the nanofabrication of superparamagnetic rods [66]. A second consequence is the observation of the sphere-to-cylinder transition for hierarchical structures. At low concentrations, the aggregates are spherical, whereas at higher mixing ratios, the structure elongate to reach size of several hundreds of nanometers. In this respect, the elongated complexes found with diblocks may appear as intermediate structures. From their sizes, they are larger than the 70 nm spherical core-shell colloids, but much smaller than the micrometric domains seen in optical microscopy (as those of Figs. 2 and 3). It is suggested here that the unidimensional growth of the structures is driven by the ability of the surfactant to self-assemble into cylindrical micelles when they are complexed with oppositely charged polyelectrolytes. It is worthwhile to mention finally that for the spherical aggregates, the micelles within the cores are disordered, as no Bragg reflection could be detected from these phase [14]. For elongated structures, the hexagonal order was systematically obtained.

### Acknowledgements :

I am grateful to Ling Qi, Jean-Paul Chapel, Pascal Hervé, Jean-Christophe Castaing from the Complex Fluids Laboratory (CRTB Rhodia, Bristol, Pa) for discussions and comments during the course of this study. I am also indebted to Annie Vacher and Marc Airiau (Centre de Recherches d'Aubervilliers, Rhodia, France) for the cryo-TEM experiments on the mixed systems. This SAXS runs were carried out at the National Synchrotron Light Source, Brookhaven National Laboratory, which is supported by the U.S. Department of Energy, Division of Materials Sciences and Division of Chemical Sciences, under Contract No. DE-AC02-98CH10886. This research is supported in part by Rhodia.